\documentclass[epj]{svjour}
\usepackage{graphics}
\begin{document}
\title{Observational constraints on decaying vacuum dark energy model}
\author{Minglei Tong\thanks{mltong@mail.ustc.edu.cn} \and Hyerim Noh% etc
\thanks{hr@kasi.re.kr}%
}                     % Do not remove
\institute{Korea Astronomy and Space Science Institute, Daejon
305-348, Korea }
\date{Received: date / Revised version: date}
\abstract{
The  decaying vacuum model (DV),
treating dark energy as a varying vacuum, has been studied well recently.
The vacuum energy decays linearly with the Hubble parameter in
 the late-times, $\rho_\Lambda(t) \propto H(t)$,
and produces the additional matter component.
 We constrain the parameters of the DV model using the recent data-sets
 from supernovae, gamma-ray bursts, baryon acoustic oscillations, CMB,
 the Hubble rate and x-rays in galaxy clusters.
  It is found that the best fit of matter density contrast $\Omega_m$ in the DV
  model is   much lager than that in $\Lambda$CDM model. We give
  the confidence contours in the $\Omega_m-h$ plane up to $3\sigma$ confidence level.
 Besides, the normalized likelihoods of $\Omega_m$ and $h$ are presented, respectively.
 % end of PACS codes
} %end of abstract
\maketitle
\section{Introduction}
\label{intro}
Since the  observations of the luminosity-redshift relation $d_L(z)$ of
type Ia supernovae (SN Ia)  indicated that the expansion of the universe
is accelerating \cite{Riess}, a cosmic component called dark energy was
 introduced to explain the acceleration within the framework of general
relativity.  Whereafter, more and more evidences, such as
cosmic microwave background (CMB)
\cite{Bennett},  large scale structure
\cite{Bahcall} especially baryon acoustic oscillations (BAO)
\cite{Eisenstein}, weak gravitational lensing \cite{Jarvis} and
x-ray clusters \cite{Allen,Allen2},  indicated that   the universe is
spatially flat and dominated by dark energy.  A
number of dark energy models have been proposed \cite{Copeland}, including scalar field
 \cite{Ratra1}, vector field \cite{Zhang94,Kiselev}, holographic dark energy
\cite{Holographic}, Chaplygin gas \cite{Kamenshchik} and so on. Among them
 the cosmological constant model ($\Lambda$CDM)
\cite{Weinberg} is the simplest one.  However, as well known, the $\Lambda$CDM suffers from the fine tuning problem: the
observed vacuum energy density of order $\sim10^{-47}$ GeV$^4$ is
about $10^{121}$ orders of magnitude smaller than the value expected
by quantum field theory for a cut-off scale being the Plank scale,
and is still about $10^{44}$ orders smaller even for a cut-off scale
being the QCD scale \cite{Copeland}.
Apart from dark energy models,  modified
gravity \cite{Carroll,Trodden} can also explain the acceleration.
However, we just focus on dark energy model in this paper.

As an extension to $\Lambda$CDM, the
the decaying vacuum (DV) dark energy model was  proposed \cite{Borges1,Carneiro}, based on the incomplete quantum field theory in the curved 4-dimension space-time. In this model, dark energy
  is described by the varying vacuum, whose energy density decays with the expansion of the universe
  leading to an additional production of the matter component. In the late-time with a quasi-de Sitter background, the vacuum density
  is proportional to the Hubble rate,
 $\rho_\Lambda(t) \propto H(t)$.  However, the equation of state  for
the vacuum is a constant value
$w=p_\Lambda(t)/\rho_\Lambda(t)=-1$, the same as that in the
$\Lambda$CDM model.
Moreover, as a interesting feature, the late-time dynamics of the DV model is
 similar to $\Lambda$CDM \cite{Borges1,Carneiro}.

  The DV model has been tested by
observational data of SN Ia \cite{Carneiro2}, and a joint data from
 SN Ia, BAO, and CMB \cite{Carneiro3,Pigozzo}. Moreover, the quasar APM 08279+5255
 at $z=3.91$
 was used to examine the DV model \cite{mltong2}, and it was found that the DV model
 can greatly alleviate the high redshift cosmic age problem existing in the $\Lambda$CDM
 model. In order to distinguish the DV model from other dark energy models at the late-time
 Universe, the statefinder and $Om$ diagnostics of the DV model were also presented in \cite{mltong2}. In order to constrain the DV model as completely as possible, besides the observational data of SN Ia of Union2 \cite{Amanullah}, BAO \cite{Percival} and  CMB \cite{Komatsu} used in the recent work \cite{Pigozzo}, we added the  data from
 Gamma-ray bursts (GRBs) at $z>1.4$ \cite{Wei},   Hubble rate \cite{Riess2,Gazta} and x-rays in galaxy clusters \cite{Allen2}.
The confidence contours in the $\Omega_m-h$ plane with the best fit values of the parameters  in the DV model will be given, using various combinations of the data. Furthermore, one-dimension  likelihoods of $\Omega_m$ and $h$, respectively, are also analyzed  with the corresponding errors up to $3\sigma$.

 In the following, we  use the  unit $8\pi G \equiv c \equiv 1$.

\section{The decaying vacuum model}
\label{sec:1}
\begin{figure}
 \resizebox{100mm}{!}{\includegraphics{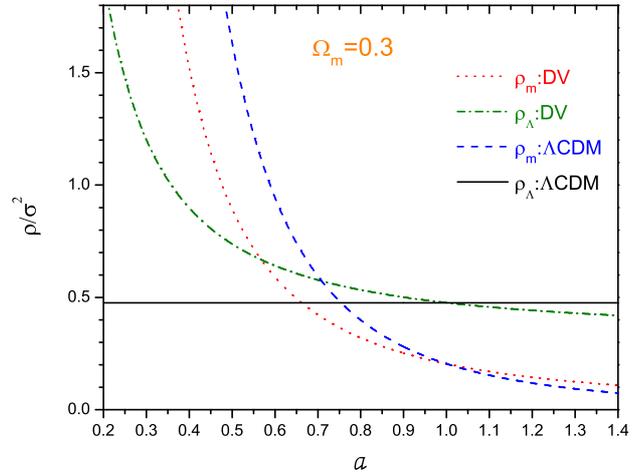}
}
 \caption{\label{rho}\small The dynamic evolutions of the energy density of matter and
 vacuum in DV model and $\Lambda$CDM model, respectively, for a fixed value of the present
 matter density contrast $\Omega_m=0.3$.
  }
 \end{figure}

In a spatially flat Friedmann-Lemaitre-Robertson-Walker (FLRW) space-time,
the Friedmann equation is
\begin{equation}\label{fridmann} \rho_T=3H^2, \end{equation}
where the total
energy density  $\rho_T=\rho_m+\rho_r+\rho_{DE}$ is composed with
matter, radiation and dark energy. $H=\dot{a}/a$ is the
Hubble rate of expansion, where an overdot means taking derivative
with respect to $t$, the cosmic time. The dynamic evolution of the total
energy density is determined by the following conservation
equation  \begin{equation}\label{totalconserve}
\dot{\rho}_T+3H(\rho_T+p_T)=0, \end{equation}
where  $p_T$
  stands for  the total pressure.
For the present epoch, the radiation component can be ignored,
and then the Universe only consists of  matter (baryons
and dark matter) and dark energy.  So
$\rho_T=\rho_m+\rho_\Lambda$ and $p_T=p_\Lambda$. In the
DV model, the vacuum density  varies with
time,  $\rho_\Lambda =\Lambda(t)$, and moreover, the covarience of the
Einstein equation
demands the equation of state $w= -1$, so  $p_\Lambda=-  \Lambda(t)$.
Thus, Eq. (\ref{totalconserve}) can
be written as
 \begin{equation} \label{matter}
\dot{\rho}_m+3H\rho_m=-\dot{\Lambda}(t),
 \end{equation}
  showing that the  decaying
vacuum density $\Lambda(t)$ plays the role of a source for matter
production. If $\Lambda$ is a  constant, Eq.(\ref{matter}) reduces
to the evolution equation for matter in $\Lambda$CDM model. To
proceed, one considers  the late-time ansatz for DV model \cite{Borges1,Carneiro}:
\begin{equation} \label{ansatz}
\Lambda=\sigma H,
\end{equation}
where $\sigma$ is a positive constant.
 According to Eqs. (\ref{fridmann}) and (\ref{ansatz}), $\sigma$ is determined as
 $\sigma = 3\Omega_\Lambda H_0$, where $\Omega_\Lambda$ is the present
value of the relative  dark energy density and  $H_0=100\, h$ km s$^{-1}$Mpc$^{-1}$
is the Hubble constant. Combining  Eq. (\ref{matter}) and Eq. (\ref{ansatz}), one can
easily get the evolution equation of the Hubble parameter
\begin{equation}
2\dot{H}+3H^2-\sigma H=0,
\end{equation}
which can be rewritten as
\begin{equation}\label{hevolution}
2H'+3H-\sigma=0,
\end{equation}
with a prime meaning  $'\equiv d/d(\ln{a})$. Eq. (\ref{hevolution}) can be  easily resolved to get
a solution of the Hubble parameter
\cite{Borges1}
\begin{equation}\label{H}
H=\frac{\sigma}{3}\left(1+\frac{\Omega_m}{\Omega_\Lambda} a^{-3/2}\right).
\end{equation}
Therefore, the matter energy density $\rho_m=3H^2-\sigma H$ and the vacuum energy density $\Lambda
=\sigma H$ have the following explicit forms
\begin{eqnarray} \label{rhomatter}
&&\rho_m(a)=\frac{\Omega_m^2\sigma^2}{3\Omega_\Lambda^2a^3}
    +\frac{\Omega_m\sigma^2}{3\Omega_\Lambda a^{{ 3}/{2}}},
\\\label{lambdamatter}
&&\Lambda(a)=\frac{\sigma^2}{3}
    +\frac{\Omega_m\sigma^2}{3\Omega_\Lambda a^{{ 3}/{2}}},
\end{eqnarray}
where we have chosen $a(t_0)=1$.
 The first terms in Eqs. (\ref{rhomatter}) and
(\ref{lambdamatter}) give the usual scaling of the matter and the
vacuum, respectively, while the second terms describe the matter
production caused by the decaying vacuum.
Eqs. (\ref{rhomatter}) and  (\ref{lambdamatter}) make  an
smooth transition between the matter  and vacuum epochs, which are illustrated in
Fig. \ref{rho} as an example with $\Omega_m=0.3$. For comparison, we also plot the cases for $\Lambda$CDM. One can see that the equal matter-vacuum dominated epoch in DV model is earlier ($\Delta z\simeq0.4$) than
that  in $\Lambda$CDM model.
 From another point of view, the evolution of the Hubble
parameter in Eq. (\ref{H}) as a function of the redshift  can be rewritten as
\begin{equation}\label{Hz}
H(z)=H_0[1-\Omega_m+\Omega_m(1+z)^{3/2}].
\end{equation}
Note that, this expression is rather different from that in the standard $\Lambda$CDM
model, due to the matter production.  In particular, if $\Lambda=0$ and $\Omega_m=1$,
one has $H(z)=H_0(1+z)^{3/2}$ leading to $\rho_m=3H_0^2(1+z)^3$, as expected
for the Einstein-de Sitter scenario. On the other hand, if $\Omega_\Lambda=1$ and
$\Omega_m=0$, we obtain $H(z)=H_0$ and $\Lambda=3H_0^2$, i.e.,
the Universe is  described by the exact de Sitter
space-time and the dark energy density will not vary.
In the following, we will discuss the constraints on
  the DV model
using various observational data.

\section{  Observational constraints }

In the following, we use the observational data from
SN Ia, BAO, CMB, GRB, Hubble parameter and x-ray in the galaxy
clusters to constrain the two free parameters $(\Omega_m, H_0)$ of
the DV model. As the likelihood function relates to $\chi^2$ function
as $\mathcal{L}\propto \exp{(-\chi^2/2)}$, one should minimize the $\chi^2$
function to obtain the best fit values of the free parameters.

\subsection{SN Ia}

As did in many literatures,  we compare the observed distance modulus $\mu_{obs}(z_i)$ of the 557 SN Ia assembled in the
Union2 compilation \cite{Amanullah}
with the theoretical distance modulus
\begin{equation}
\mu_{th}(z|p_s)=m-M=5\log_{10}\frac{d_L}{Mpc}+25,
\end{equation}
where $m$ and $M$ are the apparent and absolute magnitudes, respectively,
the complete set of parameters is $p_s\equiv (\Omega_m, H_0)$,
 and
\begin{equation}
 d_L(z)=(1+z)\int_0^z \frac{dz'}{H(z')}
 \end{equation}
  is the
luminosity distance in a spatially flat Universe.
For the  SN Ia data, the best fit to the set of
parameters $p_s$ can be obtained by minimizing
the  $\chi^2$ function
\begin{equation}\label{chi2SN}
\chi^2_{SN}(p_s)
 =\sum_{i=1}^{557}\frac{[\mu_{th}(z_i)-\mu_{obs}(z_i)]^2}{\sigma_i^2},
\end{equation}
where $\sigma_i$ stands for the $1\sigma$ uncertainty associated with the
$i$th data point.

\subsection{GRB}

Gamma-ray bursts (GRB) have been proposed to be a complementary
probe to SN Ia \cite{Schaefer}. Their high energy photons
in the gamma-ray band are most intense explosions immune
to dust extinction compared to supernovae. So far, there
are many GRB observed in the range of $0.1<z\leq8.1$,
and the maximum redshift of GRB is expected to be 10 or even larger
\cite{Bromm}. However, there is a circularity problem in the direct
use of GRB \cite{Ghirlanda}, mainly due to the lack of a set
of low-redshift GRB at $z<0.1$ which is cosmology-independent.
Not only some statistical methods have been proposed
to alleviate the circularity problem \cite{Ghirlanda2}, but also
the cosmology-independent method needing a large amount
of SN Ia at $z<1.4$ to calibrate
GRB was proposed in \cite{Liang}. Due to both the significant
improvements in GRB and SN Ia, the calibration of GRB
was updated \cite{Wei},
which can be used to constrain cosmological models without
the circularity problem. The 59 calibrated high-redshift ($z>1.4$) GRB were obtained
with the observed redshift $z$ and  modulus $\mu$
is shown in Table 2 of Ref. \cite{Wei}. Hence, the Eq. (\ref{chi2SN})
is also valid for GRB,

\begin{equation}\label{chi2GRB}
\chi^2_{GRB}(p_s)
 =\sum_{i=1}^{59}\frac{[\mu_{th}(z_i)-\mu_{obs}(z_i)]^2}{\sigma_i^2}.
\end{equation}

\subsection{BAO}

In the
distribution of SDSS luminous red galaxies \cite{sdss,Eisenstein} the distance parameter $A$ for the measurement of the BAO peak
has been usually used as a kind of examination in literatures. The definition
of $A$ is
\begin{equation}
A\equiv D_V\frac{\sqrt{\Omega_mH^2_0}}{z},
\end{equation}
where $D_V$ is the dilation scale, defined as
\begin{equation}
D_V(z)=\left[\frac{z}{H(z)}\left(\int_0^{z}
\frac{dz'}{H(z')}\right)^2\right]^{1/3}.
\end{equation}
However, as pointed out in \cite{Carneiro3}, the distance parameter $A$
is not appropriate to test the DV model, since it is obtained from the data
in the context of a $\Lambda$CDM model  and can
be considered as a good approximation only for some class of dark energy model
\cite{Doran}. We  follow Carneiro et al. \cite{Carneiro3},
using $D_V$ instead which is weakly sensitive to the cosmological
evolution before $z=0.35$.
 The  dilation scale
 $D_V$ has been observed by the SDSS at $z=0.35$ \cite{Eisenstein}, as well
as by the 2dFGRS at $z=0.2$ \cite{Percival}.
Moreover, the measurements  of SDSS and 2dFGRS
yield the ratio $D_V(0.35)/D_V(0.2)=1.736\pm0.065$ \cite{Percival}.
Then the best fit
values for the  model can be obtained by minimizing \cite{Duran}
\begin{equation}
\chi_{BAO}^2(p_s)=\frac{([D_v(0.35)/D_v(0.2)]_{th}-
[D_v(0.35)/D_v(0.2)]_{obs})^2}{\sigma^2_{D_v(0.35)/D_v(0.2)}},
\end{equation}
where $\sigma^2_{D_v(0.35)/D_v(0.2)}=0.065$.  However, as presented in Ref. \cite{Pigozzo},
it does not impose any constraint on $h$.

\subsection{CMB}
The CMB shift parameter defined as \cite{Bond,WangYu}
\begin{equation}
R\equiv\sqrt{\Omega_m}H_0\int_0^{z_{rec}}\frac{dz}{H(z)},
\end{equation}
with $z_{rec}$ the redshift of recombination,
often serves as a contrast in a large class
of dark energy models to be compared with  that
in the $\Lambda$CDM model. However, this is only
valid if the acoustic horizon at the time of last scattering
is the same \cite{Elgaroy}. This is not true
in DV model, as pointed in \cite{Carneiro3}, because for the
same values of $H_0$ and $\Omega_m$ we have different
expressions for cosmological parameters at high redshifts, due to
the process of matter production.
In this paper,  to   employ less  fitting expressions,  we use the acoustic scale $l_A$ instead of the
position for the first peak $l_1$ in the spectrum of CMB anisotropies \cite{Carneiro3,Pigozzo}.
  The acoustic scale at decoupling epoch
is defined as
\begin{equation}
l_A=\frac{\pi \int_0^{z_{*}}\frac{dz}{H(z)}}{\int_{z_{*}}^\infty
\frac{c_s dz}{H(z)}},
\end{equation}
where $z_{*}$ is the redshift of decoupling epoch,
and
\begin{equation}
c_s=\left(3+\frac{9}{4}\frac{\Omega_b}{\Omega_\gamma (1+z)}\right)^{-1/2}
\end{equation}
is the sound speed of the photon-baryon fluid.
Here $\Omega_b$ and $\Omega_\gamma$ are the present
relative energy densities of baryons and photons.
Below, we follow the observational results in \cite{Komatsu}:
$l_{A,obs}=302.09$, $z_{*}=1091$, $\Omega_b=0.0455$ and $\Omega_\gamma=2.469\times10^{-5}h^{-2}$
for $T_{CMB}=2.725K$.
Moreover, at  the very early stage for the DV model, the appropriate
generalization of Eq. (\ref{Hz}) including radiation is given by
\cite{Carneiro3}
\begin{equation}
\frac{H(z)}{H_0}\approx \{[1-\Omega_m+\Omega_m(1+z)^{3/2}]^2+\Omega_r(1+z)^4\}^{1/2},
\end{equation}
where $\Omega_r$ is the present relative energy density of
radiation. Here,  $\Omega_r=\Omega_\gamma(1+0.227\times N_{eff})$ with $\Omega_\gamma$  the relative energy density of photons and $N_{eff}=3.04$ the effective number of the standard neutrino species \cite{Mangano}.
Therefore, the corresponding $\chi^2$ function is
given by
\begin{equation}
\chi_{CMB}^2(p_s)=\frac{(l_{A,th}-l_{A,obs})^2}{\sigma_{l_{A}}^2},
\end{equation}
with $\sigma_{l_{A}}=0.76$ \cite{Komatsu}.

\subsection{Hubble rate}

Recently, some high precision
measurements constrained $H(z)$ at $z=0$ from the observation of 240 Cepheid variables
of rather similar periods and metallicities \cite{Riess2}.
The Hubble parameter as a function of redshift can be written as
\begin{equation}
H(z)=-\frac{1}{1+z}\frac{dz}{dt}.
\end{equation}
Then $H(z)$ is obtained once $dz/dt$ is known.
Simon {\it et al.} \cite{Simon} and Stern
{\it et al.} \cite{Stern} obtained $H(z)$ in the range of $0\leq z \leq1.8$,
using the differential ages of passively-evolving galaxies and archival data.
Besides,  $H(z)$ at
$z=0.24$, $0.34$ and $0.43$ is obtained \cite{Gazta} by using the BAO peak position as
a standard ruler in the radial direction. We employ the twelve data in \cite{Riess2,Stern}
and the three data in \cite{Gazta}.
The best fit values of the model parameters from observational Hubble data are determined
by minimizing
\begin{equation}
\chi^2_{Hub}(p_s)=\sum_{i=1}^{15}\frac{[H_{th}(z_i)-H_{obs}(z_i)]^2}{\sigma^2(z_i)}.
\end{equation}

%\begin{figure}[t!]
% \resizebox{170mm}{!}{\includegraphics{fig3.eps}
% \includegraphics{fig4.eps}}
% \caption{\label{fig3}
%aa}
%\end{figure}
\subsection{X-rays in galaxy clusters}

The baryons in clusters of galaxies are in the form
of hot x-ray emitting gas clouds. Thus, the fraction of
baryons in clusters, $f_{gas}$, is defined as the ratio of the x-ray
gas mass to the total mass of a gas mass fraction,
which is a constant and independent of redshift \cite{White}.
Here, we use 42 Chandra measurements of relaxed galaxy clusters in the
redshift range $0.05<z<0.1$ \cite{Allen2}.
To fit the data one can employ the empirical formula
\cite{Duran}
\begin{equation}\label{fgas}
f_{gas}(z)=\frac{K A \gamma b(z)}{1+s(z)}\left(\frac{\Omega_{b}}{\Omega_m}\right)
\left(\frac{d_A^{\Lambda CDM}}{d_A}\right)^{3/2},
\end{equation}
where the angular correction factor $A$ is   approximately
given by
\begin{equation}
A\approx\left(\frac{H(z)d_A(z)}{[H(z)d_A(z)]^{\Lambda CDM}}\right)^\eta.
\end{equation}
Here $d_A$ is the angular diameter distance, being related with the luminosity distance by
\begin{equation}
d_A(z)=\frac{d_L(z)}{(1+z)^2},
\end{equation}
and the index $\eta=0.214\pm0.022$ \cite{Allen2}. In equation (\ref{fgas}),
$K$ describes the combined effects of the residual uncertainties such as the instrumental calibration
and certain x-ray modelling issues. The parameter $\gamma$ denotes permissible departures from the
assumption of hydrostatic equilibrium due to non-thermal pressure support. The bias $b(z)=b_0(1+\alpha_b z)$ accounts for uncertainties in the cluster depletion factor, and $s(z)s_0(1+\alpha_s z)$ answers for uncertainties of the baryonic mass fraction in stars.
Most of these  parameters have large  uncertainties which can be seen in Table 4 in \cite{Allen2}.
For simplicity, we fix these parameters in the following: $K=\gamma=1$, $s_0=0.16h^{1/2}$, $\alpha_s=0$,
$b_0=0.835$ \cite{Eke}, and $\alpha_b=0$.
The $\chi^2$ function from x-rays in galaxy clusters is then given by
\begin{equation}\label{chi2x}
\chi^2_{x-rays}(p_s)=\sum_{i=1}^{42}\frac{([f_{gas}(z_i)]_{th}-[f_{gas}(z_i)]_{obs})^2}{\sigma^2(z_i)}.
\end{equation}

\section{Results and discussions}

\begin{figure*}
\begin{center}
\resizebox{160mm}{!}{\includegraphics{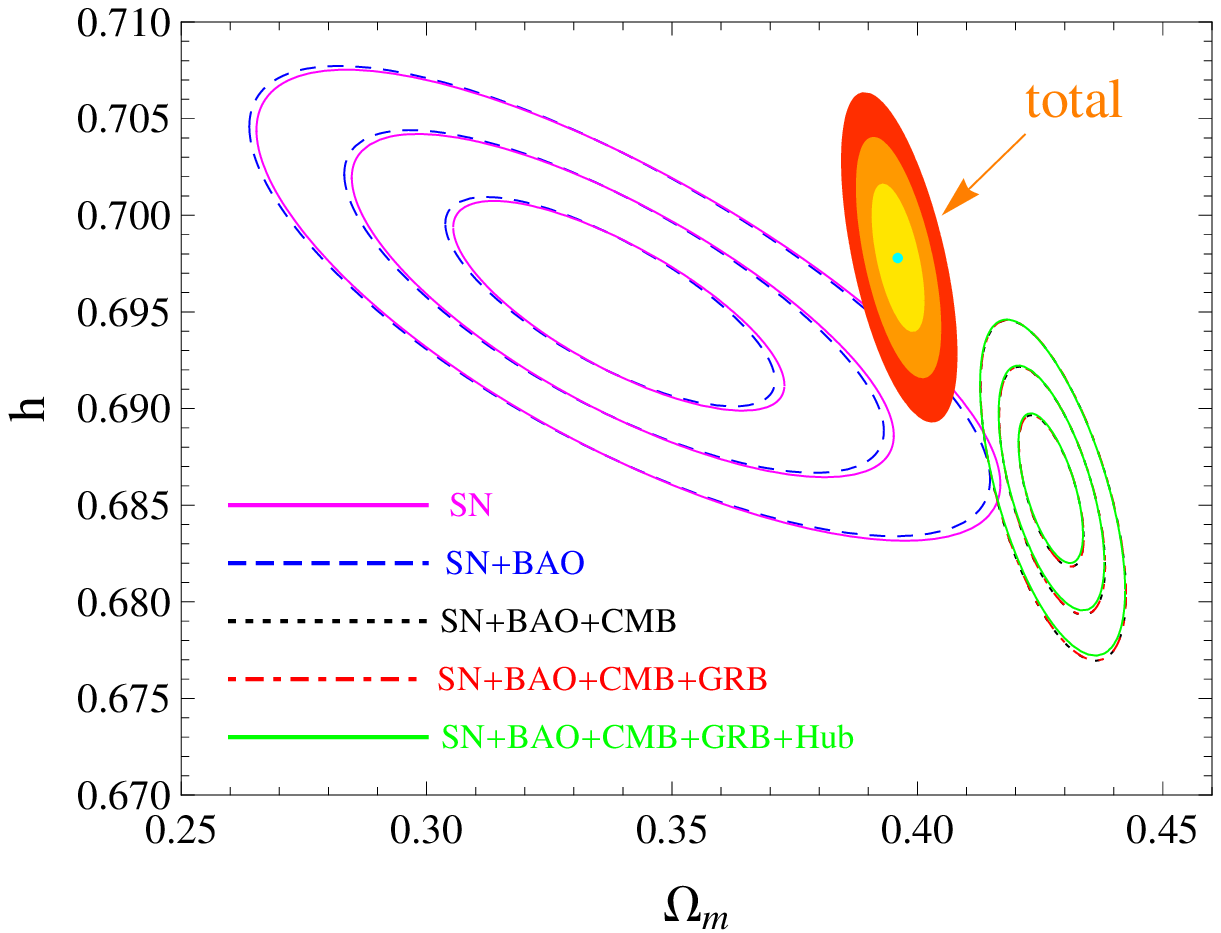}
 \qquad  \includegraphics{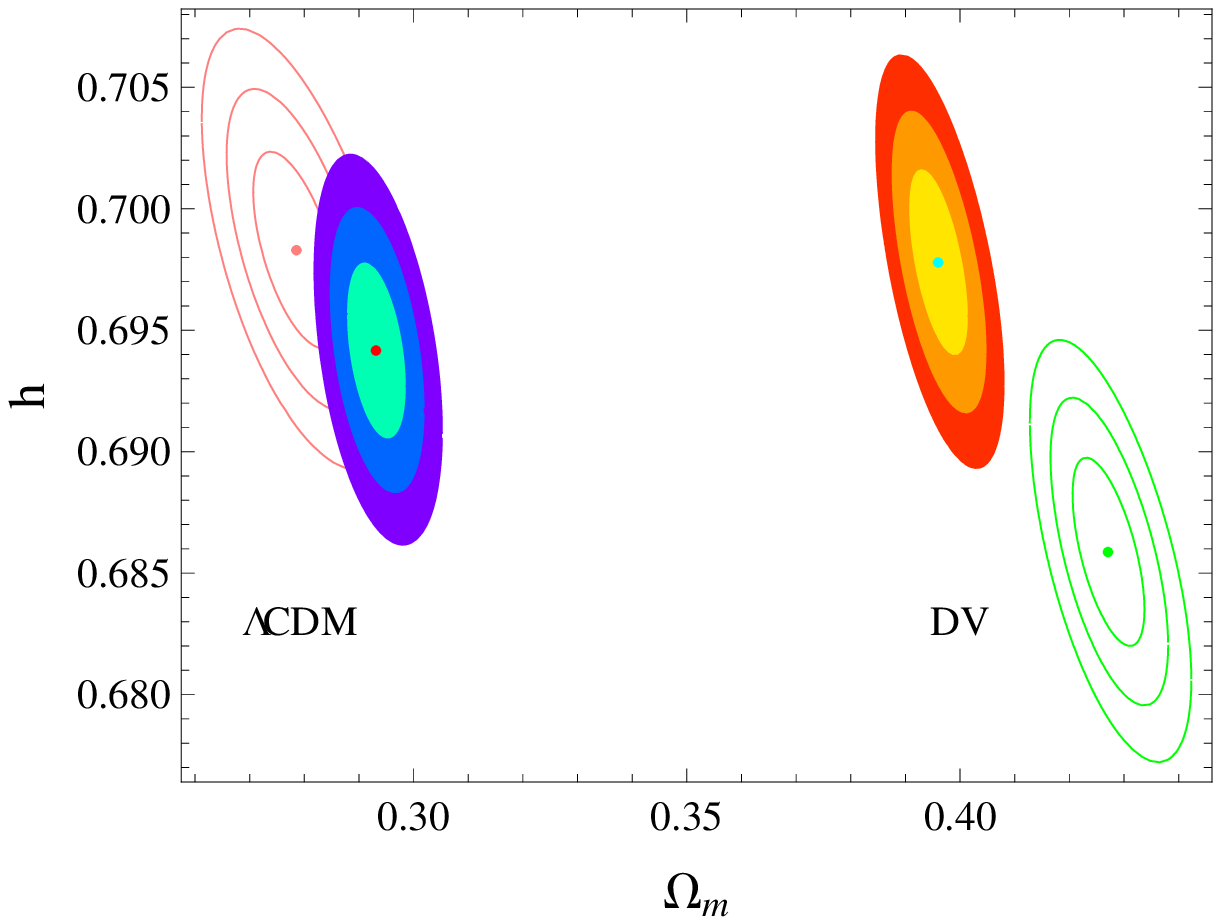}}
\vspace*{0.5cm}       % Give the correct figure height in cm
\caption{Left panel: The confidence contours up to $3\sigma$ in $\Omega_m-h$ plane
 for various combinations of observational data. The constraints from total data
 are shown as shaded contours.  Right panel: the  constraints for the DV model and $\Lambda$CDM model
 from the total data and the combined data without x-rays, respectively.
 The  solid points stand for the location of the best
 fit values.}
\label{allcontour}       % Give a unique label
\end{center}
\end{figure*}
Based on the discussion above,   the total $\chi^2$ is  combined   as
\begin{equation}\label{chi2totoal}
\chi^2_{total}=\chi_{SN}^2+\chi_{GRB}^2+\chi_{BAO}^2+\chi_{CMB}^2+\chi_{Hub}^2+\chi_{x-rays}^2.
\end{equation}
 As the likelihood function is determined as
 $\mathcal{L}\propto {\rm exp}(-\chi^2/2)$, the left panel of Fig.\ref{allcontour} shows the $68.3\%$ ($1\sigma$), $95.4\%$ ($2\sigma$) and $99.7\%$ ($3\sigma$) confidence contours of the parameters ($\Omega_m,h$) obtained by various combinations of observational data-sets for the DV model. In order to find how much a set of data
 contributes to the constraints, we added the sets of data one by one.
 We can see that the data from CMB and x-rays give very sensitive contributions to the constraints.
 Here we should emphasize two aspects. Firstly, as discussed in \cite{Carneiro3}, we have used parameters
that are strictly correct for the $\Lambda$CDM case, which would lead to bias in our results in spite of a good approximation. So,  more complete analyses of CMB need to be done.
Secondly,  as can be seen in Eq. (\ref{fgas}), the value of  $\chi^2_{x-rays}$ depends on
the ratio of the luminosity distances predicted by $\Lambda$CDM and ESF, which will deviate from 1 more and more
at higher redshift. This is why the result  of the best fits is so different with and without
 the x-rays data. Furthermore, there are many uncertain parameters included in Eq. (\ref{fgas}). Different choices of the parameters   lead to different contributions from x-ray data to the results.
On the other hand, the additions of the data from  GRB and Hubble parameter give  negligible  influences on the constraints. One can see from Fig. \ref{allcontour} that, the red dash-dotted
contours and the black dotted contours overlap with each other, and the green solid contours almost
overlap with the previous two sets of contours.
Concretely,   the best fit values of parameters $\Omega_m$ and $h$ with corresponding $\chi^2_{min}$ and $\chi^2_{min}/{dof}$ ($dof=$ degrees of freedom)  obtained from
  different combinations of data-sets are presented in Table \ref{table}, respectively. In Table \ref{table}, we can see that the best fit values of $\Omega_m$ and $h$  are very sensitive to the
  data from CMB and x-rays, but are very insensitive to the data from GRB and Hubble parameter. For comparison, we also give the two cases in the $\Lambda$CDM model.
 A comparison of confidence contours  between DV and $\Lambda$CDM can be seen in the right panel of Fig. \ref{allcontour}. The lined contours are obtained from joint  data of SN+BAO+CMB+GRB+Hub, and
  shaded contours are obtained from the total observational data. The results of the confidence contours obtained without x-rays data, as can be seen in the right panel of Fig. \ref{allcontour}
  are in good agreement with those appeared in \cite{Carneiro3,Pigozzo}.

\begin{table*}
\caption{The best fit values of the DV model using different combinations of data-sets.}
\label{table}       % Give a unique label
\begin{center}
\begin{tabular}{l|cccc}
\hline\noalign{\smallskip}
Data & $\Omega_m$ &\quad $h$ &\quad $\chi^2_{min}$ & $\chi^2_{min}/dof$  \\
\noalign{\smallskip}\hline\noalign{\smallskip}
SN  & $0.3388$ & $\quad 0.6953$ &\quad $544.587$ & $0.98$\\
SN+BAO &0.3369  &\quad 0.6955 &\quad 545.758 &0.98\\
SN+BAO+CMB  & 0.4272 &\quad 0.6857 &\quad 561.724&1.01 \\
SN+BAO+CMB+GRB & 0.4272 &\quad 0.6857 &\quad 584.866 & 0.95\\
SN+BAO+CMB+GRB+Hub & 0.4271 &\quad 0.6859 &\quad 596.345 & 0.95\\
total & 0.3960 & \quad 0.6978 & \quad 966.343 & 1.44\\
SN+BAO+CMB+GRB+Hub ($\Lambda$CDM) & 0.2785 &\quad 0.6983 &\quad 577.915 & 0.92\\
total ($\Lambda$CDM) & 0.2931 & \quad 0.6941 & \quad 632.412 & 0.94\\
\noalign{\smallskip}\hline
\end{tabular}
% Or use
\vspace*{0.06cm}  % with the correct table height
\end{center}
\end{table*}

\begin{figure*}
\begin{center}
\resizebox{160mm}{!}{\includegraphics{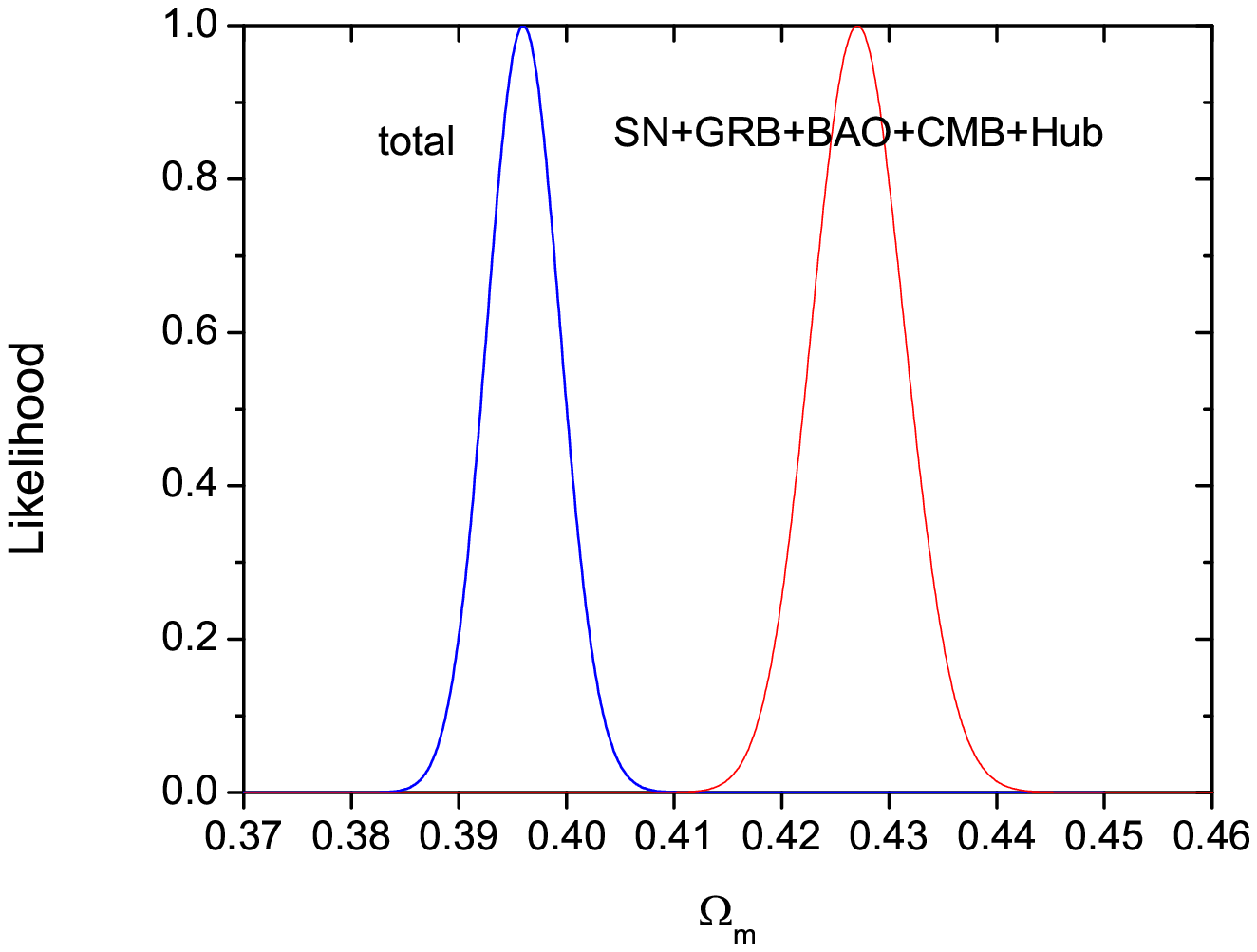}
  \includegraphics{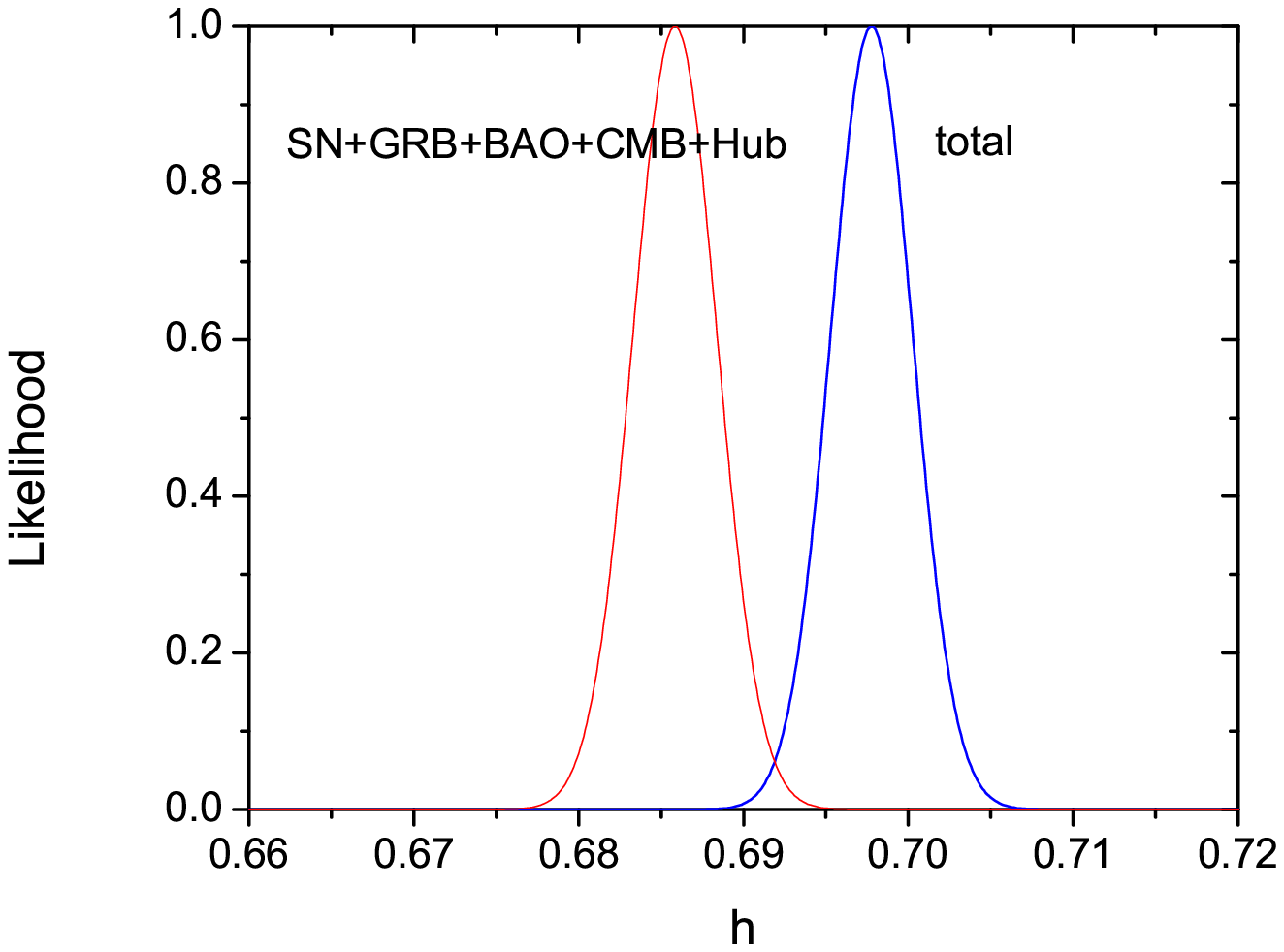}}
\vspace*{0.5cm}       % Give the correct figure height in cm
\caption{The normalized likelihoods of the parameters $\Omega_m$ and $h$, respectively,
 for the DV model.}
\label{onedimension}       % Give a unique label
\end{center}
\end{figure*}

Fig. \ref{onedimension} presents the 1-dimension  normalized likelihoods of the two parameters $\Omega_m$
and $h$ for DV model based on the total data (blue lines)  and the combined  data  without x-rays (red lines), respectively. Concretely, the best fit values   are $\Omega_m=0.3960^{+0.0034}_{-0.0033}(1 \sigma)^{+0.0070}_{-0.0067}
(2 \sigma)^{+0.0104}_{-0.0099} (3 \sigma)$ and $h=0.6978\pm0.0025\pm0.0050\pm0.0074$ with $\chi^2_{total}=966.343$.
If we abandon the data from x-ray,  the best fit values   are $\Omega_m=0.4270^{+0.0044+0.0089+0.0132}_{-0.0043-0.0085-0.0125}$ and $h=0.6858\pm0.0025\pm0.0050^{+0.0076}_{-0.0075}$ with $\chi^2_{total}=596.345$. One can see the big different results of the both likelihoods of $\Omega_m$ and $h$. As
discussed above, the differences are due to the deviations of the luminosity distance given by ESF from that given by
$\Lambda$CDM.

Just as the results from the examination using energy density perturbations \cite{Borges2}, the best fits  of $\Omega_m$ for DV model  are
 very large compared to those for  $\Lambda$CDM model. The data from CMB give a magnificent response
for the large value of $\Omega_m$. However, more complete analyses of CMB need to be done.  Moreover, there are
many  parameters with  large uncertainties in Eq. (\ref{fgas}), when we evaluate  $\chi^2_{x-rays}$  in
Eq. (\ref{chi2x}). This defect can just explain the large value of $\chi^2_{min}/dof=1.44$, which  can be seen in Table \ref{table}.


%

\section{Conclusions}

We constrained the parameters of the DV model using the most recent joint data from
557 SN Ia in Union2 compilation, 59 GRBs in the high redshifts ($z>1.4$), the ratio of the dilation
scale at $z=0.35$ and $z=0.2$ of BAO, the acoustic scale at decoupling epoch of CMB measured by WMAP 7, 15 data of Hubble rate, and 42 x-ray gas mass ratio  in galaxy clusters. Due to the joint data, the parameters of the best fit values are: $\Omega_m=0.3960^{+0.0034+0.0070+0.0104}_{-0.0033-0.0067-0.0099}$ and $h=0.6978\pm0.0025\pm0.0050\pm0.0074$. With the same joint data, we also calculated the best fits for $\Lambda$CDM model:  $\Omega_m=0.2931^{+0.0034+0.0069+0.0104}_{-0.0034-0.0067-0.0099}$ and $h=0.6941\pm0.0024\pm0.0047\pm0.0070$.
However, the data from x-rays depend on many fitting parameters with a large uncertainty.
Then, if  we neglect the poor data of x-rays, the results are: $\Omega_m=0.4270^{+0.0044+0.0089+0.0132}_{-0.0043-0.0085-0.0125}$ and $h=0.6858\pm0.0025\pm0.0050^{+0.0076}_{-0.0075}$ for the DV model; $\Omega_m=0.2785^{+0.0054+0.0109+0.0164}_{-0.0052-0.0103-0.0151}$ and $h=0.6983\pm0.0027\pm0.0053\pm0.0079$ for the $\Lambda$CDM model.
The best fit value of $\Omega_m$ in the DV model is larger than that in the $\Lambda$CDM model, since there is additional production of the matter component due to the decaying of the vacuum energy. However, we cannot rule out the DV model from the current observational data. We should wait for
more abundant, accurate, and model-independent data to constrain this model. On the other hand, the DV model alleviates the fine tuning problem and high redshift cosmic age problem which exist in $\Lambda$CDM model.

\

{ACKNOWLEDGMENT}: H. Noh's  work has been  supported by Mid-career Research Program through
 National Research Foundation funded by the MEST (No. 2010-0000302).

%% BibTeX users please use
%% \bibliographystyle{}
%% \bibliography{}

\small
\begin{thebibliography}{88}

\bibitem{Riess} A.G. Riess   {et al}., Astron. J. {\bf 116},  1009 (1998);

                A.G. Riess   {et al}., Astrophys. J. {\bf 117}, 707 (1999);

                S. Perlmutter   {et al}., Astrophys. J. {\bf517},  565 (1999).




\bibitem{Bennett} C.L. Bennett   {et al}., Astrophys. J. Suppl.{\bf 148},  1 (2003);

                  D.N. Spergel   {et al}., Astrophys. J. Suppl. {\bf148},  175 (2003);

                  H. V. Peiris   {et al}., Astrophys. J. Suppl. {\bf148}, 213 (2003);

 D.N. Spergel   {et al}., Astrophys. J. Suppl. {\bf170}, 377 (2007);

              L. Page   {et al}., Astrophys. J. Suppl. {\bf170}, 335 (2007);

              G. Hinshaw   {et al}., Astrophys. J. Suppl. {\bf170}, 263 (2007);



 G. Hinshaw     {et al}., Astrophys. J. Suppl. {\bf180}, 225 (2009);

  M.R. Nolta   {et al}., Astrophys. J. Suppl. {\bf180}, 296 (2009);

  J. Dunkley    {et al}., Astrophys. J. Suppl. {\bf180}, 306 (2009);

  E. Komatsu   {et al}., Astrophys. J. Suppl. {\bf180}, 330 (2009).




\bibitem{Bahcall} N.A. Bahcall, J.P. Ostriker, S. Perlmutter, P.J. Steinhardt,
                  Science {\bf284}, 1481 (1999).

\bibitem{Eisenstein}  D.J. Eisenstein  et al, Astrophys. J. {\bf 633}, 560 (2005).

\bibitem{Jarvis} J. Jarvis, B. Jain, G. Berstein, D. Dolney,
                            Astrophys. J. {\bf 644}, 71 (2006);

                  H. Heokstra   {et al}.,  Astrophys. J. {\bf 647}, 116 (2006);

                  R. Massey   {et al}., Nature   {\bf445},  286 (2007).

\bibitem{Allen} S.W. Allen   {et al}., Mon. Not. Roy. Astron. Soc. {\bf353}, 457 (2004).

\bibitem{Allen2} S.W. Allen   {et al}., Mon. Not. Roy. Astron. Soc. {\bf 383},  879 (2008).
\bibitem{Copeland} E.J. Copeland,   M. Sami,    S. Tsujikawa,
                    {Int. J. Mod. Phys.} D  {\bf15}   1753 (2006).
\bibitem{Ratra1} B. Ratra, P.J.E Peebles,
{Phys. Rev.}  D  {\bf37}  3406 (1988);

 I. Zlatev,  L.M Wang,    P.J. Steinhardt,
 {Phys. Rev. Lett.} {\bf82},  896 (1999);

  P.J. Steinhardt, L. Wang,
  I. Zlatev,  {Phys. Rev.} D  {\bf59}, 123504 (1999);

     P.G. Ferreira, M. Joyce,  {Phys. Rev.} D   {\bf58},  023503 (1998);

  S. Dodelson, M. Kaplinghat,      E. Steinhart,
               {Phys. Rev. Lett.} {\bf85},  5276 (2000);

 F.C. Carvalho,  J.S. Alcaniz, J.A.S. Lima,    R. Silva,    {Phys. Rev. Lett.} {\bf97},  081301 (2006);
 
 M.L. Tong, Y. Zhang, Z.W. Fu, Class. Quantum Grav. {\bf28}, 055006 (2011).

  \bibitem{Zhang94}   Y. Zhang,   {Phys. Lett.} B   {\bf340}  18 (1994);

 Y. Zhang, {Gen. Relativ. Gravit.} {\bf34},  2155  (2002);

Y. Zhang, Gen. Relativ. Gravit. {\bf35}, 689 (2003);

W. Zhao, Y. Zhang,  Phys. Lett. B {\bf690}, 64 (2006);

Y. Zhang, T.Y. Xia,    W. Zhao,
                  Class. Quantum Grav. {\bf 24},  3309 (2007);

 T.Y. Xia, Y. Zhang, Phys. Lett. B {\bf 656}, 19 (2007);

   M.L. Tong, Y. Zhang, T.Y. Xia,
             Int. J. Mod. Phys. D {\bf 18},  797 (2009).
\bibitem{Kiselev} V.V. Kiselev,   {Class. Quantum Grav.} {\bf21},  3323 (2004);

  C.    Armendariz-Picon,  {JCAP}  {\bf07}, 007 (2004).






\bibitem{Holographic}   A.G. Cohen,  D.B. Kaplan,   A.E. Nelson,  {Phys. Rev. Lett.} {\bf82},  4971 (1999);

  D. Pav\'{o}n,      W. Zimdahl,  {Phys. Lett.} B  {\bf628},  206 (2005);

   M. Li,  {Phys. Lett.} B, {\bf603}   1 (2004).

   C. Gao, F. Wu,     X. Chen,  {Phys. Rev.} D  {\bf79},  043511 (2009).


\bibitem{Kamenshchik} A.Y. Kamenshchik, U. Moschella,   V. Pasquier, Phys. Lett. B, {\bf511},  265(2001);

M.C. Bento, O. Bertolami,   A.A. Sen, Phys. Rev. D {\bf66},  043507 (2002).

\bibitem{Weinberg} S. Weinberg, Rev. Mod. Phys. {\bf61}, 1 (1989);

                   T. Padmanabhan, Phys. Rep. {\bf380}, 235 (2003).

\bibitem{Carroll} L.Parker,   A. Raval, Phys. Rev. D {\bf60}, 063512 (1999);
                                         Phys. Rev. D {\bf60}, 123502 (1999);

              S.M. Carroll, V. Duvvuri, M. Trodden, M.S. Turner
                                          Phys. Rev. D {\bf70}, 043528 (2004).

\bibitem{Trodden} M. Trodden, arXiv:1011.0861;

                  V.T. Toth. arXiv:1011.5174.
\bibitem{Borges1} H.A. Borges, S. Carneiro,
               Gen. Relativ. Gravit. {\bf37},    1385   (2005).

\bibitem{Carneiro} S. Carneiro, J. Phys. A {\bf40}, 6841 (2007).



\bibitem{Carneiro2} S. Carneiro,  C. Pigozzo, H.A. Borges,   J.S. Alcaniz,
                   Phys. Rev. D {\bf74}, 023532 (2006).

\bibitem{Carneiro3} S. Carneiro, M.A. Dantas, C. Pigozzo,   J.S. Alcaniz,
                   Phys. Rev. D {\bf 77}, 083504 (2008).
\bibitem{Pigozzo} C. Pigozzo, M. A. Dantas, S. Carneiro,   J. S. Alcaniz, arXiv:1007.5290.

\bibitem{mltong2} M.L. Tong, Y. Zhang, Phys. Rev. D  {\bf80}, 023503.
\bibitem{Amanullah} R. Amanullah,  {et al}.,
Astrophys. J. {\bf716}, 712 (2010).
\bibitem{Wei} H. Wei, JCAP {\bf08},  020 (2010).

\bibitem{Percival} W.J. Percival   {et al}., Mon. Not. R. Astron. Soc. {\bf401}, 2148 (2010).


\bibitem{Komatsu} Komatsu   {et al}., arXiv:1001.4538.
\bibitem{Riess2} A.G. Riess   {et al}.,  Astrophys. J. {\bf659}, 98 (2007).

\bibitem{Gazta} E. Gazta\~{n}aga, A. Cabr\'{e},  L. Hui, Mon. Not. R. Astron. Soc. {\bf399}, 1663 (2009).


\bibitem{Schaefer} B.E. Schaefer, Astrophys. J. 660, 16 (2007).

  \bibitem{Bromm} V. Bromm,   A. Loeb, Astrophys. J. 575, 111 (2002);

J.-R. Lin, S.N. Zhang,   T.P. Li, Astrophys. J. 605, 819 (2004).
\bibitem{Ghirlanda} G. Ghirlanda, G. Ghisellini,  C. Firmani,
New. J. Phys. 8, 123, (2008).
\bibitem{Ghirlanda2} G. Ghirlanda, G. ghisellini, D. lazzati,
  C. firmani, Astrophys. J. 613, L13, (2004);

C. Firmani, G. ghisellini, G. Ghirlanda,  V. Avila-Reese,
Mon. Not. Roy. Astron. Soc. 360, L1, (2005);

H. Li et al., Astrophys. J. 680, 92 (2008);

E.-W. Liang,   B. Zhang, Mon. Not. Roy. Astron. Soc. Lett. 369, L37 (2006).

\bibitem{Liang} N. Liang, W.K. Xiao, Y. Liu,  S.N. Zhang,
Astrophys. J. 685, 354 (2008).

\bibitem{sdss} M. Tegmark et al., Phys. Rev. D {\bf69}, 103501, (2004).

\bibitem{Doran} M. Doran, S. Stern,   E. Thommes, JCAP {\bf04}, 015 (2007).

\bibitem{Duran} I. Dur\'{a}n, D. Pav\'{o}n,   W. Zimdahl, JCAP {\bf07}, 018 (2010);

L.X. Xu,   J.B. Lu, JCAP {\bf03}, 025 (2010).



\bibitem{Bond} J.R. Bond, G. Efstathoiu,   M. Tegmark,
Mon. Not. Roy. Astron. Soc. {\bf291}, L33 (1997).


\bibitem{WangYu} Y. Wang,   P. Mukherjee, Astrophys. J. {\bf650}, 1 (2006).



\bibitem{Elgaroy} O. Elgaroy,   T. Multamaki, Astron. Astrophys. {\bf471}, 65 (2007).

\bibitem{Mangano} G. Mangano,  {et al.,}  Nucl. Phys. B {\bf729}, 221 (2005).
 \bibitem{Simon} J. Simon, L. Verde,   R. Jim\'{e}nez, Phys. Rev. D {\bf71}, 123001 (2005).
  \bibitem{Stern} D. Stern, R. Jim\'{e}nez, L. Verde, M. Kamionkowski,   S.A. Stanford,
JCAP {\bf02}, 008 (2010).
\bibitem{White} S.M.D. White, J.F. Navarro, A. Evrard,
  C.S. Frenk, Nature {\bf366}, 429 (1993).

    \bibitem{Eke}V.R. Eke, J.F. Navarro,  C.S. Frenk,  Astrophys. J. {\bf503}, 569 (1998);

             R.A. Crain,  {et al.,} Mon. Not. Roy. Astron. Soc. {\bf377}, 41 (2007).


\bibitem{Borges2} H.A. Borges, S. Carneiro,    J.C. Fabris,
Phys. Rev. D {\bf78}, 123522 (2008).
\end{thebibliography}
%%
%% Non-BibTeX users please use
%\begin{thebibliography}{}
%%
%% and use \bibitem to create references.
%Author, \textit{Book title} (Publisher, place year) page numbers
%\end{thebibliography}

\end{document}